\documentclass[aps,prd,nofootinbib,twocolumn,superscriptaddress,longbibliography,showpacs]{revtex4-1}

\usepackage{amssymb,amstext,amsmath,amsthm}
\usepackage{graphicx}
\usepackage{latexsym}
\usepackage{amsfonts}
\usepackage{verbatim}
\usepackage{color}  
\usepackage{amsmath,amsxtra,amsthm,amssymb,xr}
\usepackage[]{mathrsfs}

\newcommand{\bea}{\begin{eqnarray}}
\newcommand{\eea}{\end{eqnarray}}
\newcommand{\nn}{\nonumber}

\newcommand{\R}{\mathbb{R}}

\newcommand{\lalg}[1]{\mathfrak{#1}}  

\newcommand{\su}{\lalg{su}}

\newcommand{\dd}{{\mathrm d}}  

\def\la{\langle}
\def\ra{\rangle}

\newcommand{\bra}[1]{\la{#1}|}
\newcommand{\ket}[1]{|{#1}\ra}

\def\d{\mathrm{d}}

\newcommand{\half}{\frac{1}{2}}

\newcommand{\Ref}[1]{(\ref{#1})}
\def\f{\frac}

\newcommand{\lp}{\ell_{\rm P}}

\newcommand{\eqa}{\begin{eqnarray}}
\newcommand{\neqa}{\end{eqnarray}}
\newcommand{\be}{\begin{equation}}
\newcommand{\ee}{\end{equation}}

\newcommand{\kin}{\mathrm{Kin}}

\def\be{\begin{eqnarray}}
\def\ee{\end{eqnarray}}
\def\bw{\begin{widetext}}
\def\ew{\end{widetext}}

\newcommand{\cc}{\mathcal C}

\newcommand{\cg}{\mathcal G}
\newcommand{\ch}{\mathcal H}

\newcommand{\cl}{\mathcal L}

\newcommand{\co}{\mathcal O}

\newcommand{\cv}{\mathcal V}

\renewcommand{\d}{\delta}
\newcommand{\eps}{\epsilon}

\renewcommand{\l}{\lambda}

\newcommand{\rmd}{\mathrm d}

\newcommand{\tr}{\mathrm{Tr}}

\newcommand{\SU}{{SU}}

\def\f{\frac}

\def\ra{\rangle}
\def\la{\langle}

\def\d{\delta}
\def\de{\Delta}
\newcommand{\p}{\partial}

\newcommand{\Hil}{\mathcal{H}}
\newcommand{\bohr}{\mathrm{Bohr}}
\newcommand{\mubar}{\bar{\mu}}
\newcommand{\pam}{\f{2}{(8\pi G)^3\gamma^4}}

\newcommand{\gr}{\mathrm{gr}}
\newcommand{\m}{\mathrm{M}}

\newcommand{\svi}[1]{(\overline{N})_{#1}}

\newcommand{\ki}{\mathrm{kin}}
\newcommand{\eff}{\mathrm{eff}}

\newcommand{\bounce}{\mathrm{bounce}}
\newcommand{\sigbar}{\bar{\Sigma}}
\newcommand{\mubarp}{\mubar'}

\begin{document}

\date\today

\title{$2+1$ dimensional loop quantum cosmology of Bianchi I models}
\author{You Ding}\email{dingyou@bjtu.edu.cn}
\affiliation{Department of Physics, Beijing Jiaotong University, Beijing 100044, China}
\author{Xiangdong Zhang}\email{Corresponding author: scxdzhang@scut.edu.cn}
\affiliation{Department of Physics, South China University of
Technology, Guangzhou 510641, China}

\begin{abstract}
 We study the anisotropic Bianchi I loop quantum cosmology in 2+1 dimensions. Both the $\mubar$ and $\mubar'$ schemes are considered in the present paper and the following expected results are established: (i) the massless scalar field again play the role of emergent time variables and serves as an internal clock; (ii) By imposing the fundamental discreteness of length operator, the total Hamiltonian constraint is obtained and gives rise the evolution as a difference equation; and (iii) the exact solutions of Friedmann equation are constructed rigorously for both classical and effective level. The investigation extends the domain of validity of loop quantum cosmology to beyond the four dimensions.

\pacs{04.60.Pp, 98.80.Qc}
\end{abstract}
\maketitle
\date{\small\today}

\section{Introduction}
Loop quantum cosmology (LQC) \cite{Bojowald,Ashtekar:2011ni},
 which applies principles of loop quantum gravity (LQG) \cite{Ashtekar:2004eh,carlo,tman,Han:2005km,Chiou:2014jwa} to cosmological settings, has been developed as a symmetry-reduced model of the full theory of LQG, to implement and test main ideas of LQG. Many obscure aspects in  LQG become transparent in LQC, since in LQC the mathematical structure is much simplifier. Particularly, the space-flat ($k=0$) Friedmann-LeMa$\mathrm{\hat{i}}$tre-Robertson-Walker (FLRW) model without cosmological constant $(\Lambda=0)$ can be solved
exactly if  the scalar field is used as an internal clock already in the classical theory, prior
to quantization, and works in a suitable representation \cite{Ashtekar:2007em}. In the exactly soluble LQC model, the big bang singularity is resolved, which is replaced by a big bounce; it is obtained the analytical expression of the upper bound of the energy density operator; furthermore,
questions regarding the behavior of fluctuations and preservation of semi-classicality across
the bounce can be answered in detail. More interestingly, these features are still valid in arbitrary spacetime dimensions \cite{Zhang:2014xqa,Zhang:2015bxa}.

If one retains the homogeneity assumption but consider a
generalization of the $k=0$ FLRW model, like,  to include a cosmological constant, spatial curvature or anisotropies, one will find the singularity is always resolved, although the exact solubility does not hold any more. The quantization of homogeneous but anisotropic models is important for three reasons. First, the Belinskii-Khalatnikov-Lifshitz (BKL) conjecture in classical general relativity
says that when sufficiently close to the space-like singularity, time derivatives overwhelm the spacial ones. Thus
the behavior of the
gravitational field as one approaches generic space-like singularities can be largely understood using homogeneous but anisotropic models. This makes the question of singularity
resolution in anisotropic models conceptually important. Second, an exciting fact of bouncing cosmological models is that the scales measured in the cosmic microwave background (CMB) can be in causal contact if the current expanding phase is preceded by a contracting one, thus opening the possibility for a replacement of the standard inflationary scenario. In order to more fully develop this scenario, one must go beyond the assumption of isotropy and/or homogeneity. Third, as a symmetry-reduced model of LQG to test main ideas of the full theory, isotropic LQC is the simplest model, and inclusion of anisotropy will provide more insights.

The loop quantization of Bianchi I model was initially studied in \cite{Bojowald:2003md} and more recently in \cite{Chiou:2006qq,Ashtekar:2009vc}. Due to
the underlying complexity of the partial difference equation obtained in \cite{Ashtekar:2009vc}, numerical evolution of physical states
in this model are yet to be performed. Nevertheless, the model proposed earlier by \cite{Chiou:2006qq} is less complex and is studied more, although this quantization suffers from fiducial cell scaling
and infrared problems \cite{Szulc:2008ar,Corichi:2009pp}.

On the other hand, the study in 2+1 dimensional gravity has a long history and fruitful results. Usually it is viewed as a simpler model of 3+1 dimensional gravity, and thus may provides useful insights to solve some complicated problems we encountered in 3+1 dimensional gravity.  In particular in the field of quantum gravity, 2+1 dimensional gravity is usually employed to test the validity of quantization procedure used in 3+1 dimensions \cite{Ashtekar:1989qd,Thiemann:1997ru}, since a fully satisfied 3+1 dimensional quantum gravity is still an open problem.  In addition, it is also significance that sometimes three dimensional gravity even offers a possibility to understand the
physical issues relevant to four dimensional gravity \cite{Ma:2001dm,Garcia:2003vy}. Moreover, the 2+1 dimensional cosmology is a rather active topic both from classical and quantum perspectives \cite{Biswas:2005zn,Garcia:2003vy,Jana:2013fga,Carlip}. Furthermore, the precursor studies on 3+1 dimensional loop quantum Bianchi I models show that the planar collapse for
Kasner-like solutions is not resolved by the quantum effect\cite{Chiou:2006qq}. The reason for this can be understood intuitively since in 3+1 dimensional case, we use the discreetness of the area operator rather than the length operator as our fundamental building blocks. Thus the vanishing behavior of the length scale factors $a_I$ remains \cite{Chiou:2006qq}. However, in 2+1 dimensions, the spectrum of length operator is fundamentally discreet, thus all the singularity appeared in 2+1 dimensional classical Bianchi I model are very likely to be resolved. Thus in this paper, we will mimic the scheme proposed in \cite{Ashtekar:2009vc}, to study the quantization of Bianchi I model in 2+1 dimensions, instead of the physical 3+1 dimensions.   

This paper is organized as follows: After a brief introduction in the beginning, we present the classical dynamics of 2+1 dimensional Bianchi I models from Hamiltonian framework in Section \ref{sec1}. With this classical Hamiltonian dynamics, in section \ref{sec2}, we construct the corresponding loop quantum cosmology of 2+1 dimensional Bianchi I models. Then in sections \ref{sec3} and \ref{sec4}, we discuss the action of Hamiltonian operator and the properties of quantum dynamics respectively. Some conclusions are presented in last section.

\section{Classical dynamics of 2+1 dimensional  Bianchi I models}\label{sec1}
\subsection{The canonical pair of 2+1 dimensional  Bianchi I models}
Let us start by summarizing the classical dynamics of 2+1 dimensional  Bianchi I models.  Our spacetime manifold $M$ is topologically $\R^3$.
We restrict ourselves to diagonal Bianchi I metrics, given in terms of the directional scale factors $a_I$ with $I=1,2$
\be\label{le}
ds^2=-N^2dt^2+a_1^2dx^2+a^2_2dy^2,
\ee
where $N$ represents the lapse.
Since we consider noncompact Bianchi I model and all fields are spatially homogeneous, we can introduce an elemental cell $\mathcal{V}$ and restrict all integrations to it \cite{Ashtekar:2003hd}. We choose elementary cell $\mathcal{V}$ as its edges lie along the coordinate axis $x,y$. We fix a fiducial flat metric ${^oq_{ab}}$ with line element $ds^2_o=dx^2+dy^2$. We denote by ${}^oq$ the determinant of this metric, by $L_I$ the lengths of the two edges of $\mathcal{V}$ as measured by ${}^oq_{ab}$, and by $V_o=L_1L_2$ the volume of $\mathcal{V}$ as measured by  ${}^oq_{ab}$. Among the fiducial co-dyads compatible with ${}^oq_{ab}$,  we select ${}^o\omega_a^i$, without generality, such that
\be
{}^o\omega_a^I=D_ax^I,  \ \mathrm{and}\
{}^o\omega_a^3=0.
\ee 

With this fiducial structure at hand, we can now introduce the phase space of 2+1 dimensional  Bianchi I models, which is reduced from the one of the full theory. In the full theory of 2+1 dimensional  LQG \cite{Thiemann:1997ru}, the phase space is spanned by a canonical pair ($A_a^i, E^a_i$), with $A_a^i$ an $\su(2)$ connection and $E^a_i=\delta_{ij}\epsilon^{ab}e_b^i$ the momentum conjugate to $A_a^i$. And the symplectic structure is given by
\be
\{A_a^i(\vec{x}),E^b_j(\vec{x'})\}=8\pi G\gamma\delta_a^b\delta^i_j\delta(\vec{x},\vec{x}').
\ee
Because of Bianchi symmetry, the connections $A_a^i$ are reduced to 2 constants $ {c}^I$, and the momenta  $E^a_i$ are reduced to 2 constants $ {p}_I$:
\begin{align}
A_a^I&=c^I(L^I)^{-1}{}^o\omega^I_a,\ \mathrm{and}\ A_a^3=0;\\
E^a_I&=p_IL_IV_o^{-1}\epsilon^{ab}\delta_{IJ}\;{}^o\omega_a^J,\ \mathrm{and}\ E^a_3=0.
\end{align}
Note that there is no summation over $I$ and the repeated upper and lower $J$ is summed.
The momentum variables $p_I$ are directly related to the scale factors $a_I$ as
\be
 {p}_1=a_1L_2,\quad  {p}_2=a_2L_1.\label{pa}
\ee
As we will see below, the connections ${c}^I$ are directly related to time derivatives of the scale factors. 
The resulting non-vanishing Poisson brackets are given by
\be
\{c^I,p_J\}=8\pi G\gamma \d^I_J,\label{cp}
\ee
where $\gamma$ is the Barbero-Immirzi parameter.

\subsection{Evolution equations}
Now we come to the constraints. Because we have restricted ourselves to diagonal metrics and fixed the internal gauge, the Gauss and the diffeomorphism constraints are identically satisfied. The Hamiltonian constraints can be obtained  by restricting the spatial integrations to the fiducial cell $\mathcal{V}$:
\begin{align}
\mathcal{C}&=\cc_{\gr}+\cc_{\m}=\int_{\cv}N(\ch_{\gr}+\ch_{\m}) d^2x,
\end{align}
where $N$ is the lapse and we will fix $N=1$ in the following for simplicity.
The gravitational and the matter parts of the constraint densities are given by
\begin{align}
\ch_{\gr}&=-\f{E^a_iE^b_j}{16\pi G\gamma^2\sqrt{q}}\epsilon^{ij}{}_{k}F_{ab}{}^k\label{hamilton_gr}\\
\ch_{\m}&=\sqrt{q}\rho_{\m}.
\end{align}
Here  $F_{ab}{}^k$ is the curvature of connection $A_a^i$, and $\rho_{\m}$ is the energy density of the matter fields. The constraint density in Eq. \Ref{hamilton_gr} is different from the one in full theory of 2+1 dimensional  LQG, by considering the flat-space property of Bianchi I models.
We consider a massless scalar field $T$ coupled to gravity, thus the Hamiltonian constraint, in terms of induced variables, is given by
 \be
\cc=-\frac{c_1c_2}{8\pi G\gamma^2}+\frac{p^2_T}{2p_1p_2},\label{hamilton_class}
\ee
where $p_T$ is the conjugate momentum of the scalar field $T$, and the matter energy density is given by $\rho_{\m}=p^2_T/2V^2$.
From the Hamiltonian constraint \Ref{hamilton_class}, we find that $p_T$ is a constant of motion and $T$ is a monotonic function of $t$. Thus scalar $T$ can be considered as the internal time.
The Poisson brackets of the matter part is given by $\{T,p_T\}=1$.
The time evolution of  $p_I$ thus yield
\begin{align}
&\dot{p}_1=\{p_1,\cc\}=-8\pi G\gamma\frac{\p\cc}{\p c_1}= \f{c_2}{\gamma}\label{dotp1}\\
&\dot{p}_2=\{p_2,\cc\}=-8\pi G\gamma\frac{\p\cc}{\p c_2}= \f{c_1}{\gamma}\label{dotp2}\\
&\dot{c}_1=\{c_1,\cc\}=8\pi G\gamma\f{\p \cc}{\p p_1}=-8\pi G\gamma p_2\rho_{\m}\\
&\dot{c}_2=\{c_2,\cc\}=8\pi G\gamma\f{\p \cc}{\p p_2}=-8\pi G\gamma p_1\rho_{\m}\label{dotc2}
\end{align}
Combining Eqs. \Ref{pa}, \Ref{dotp1} and \Ref{dotp2}, one gets
\be
{c}_I= \gamma \dot{a}_JL_I,\label{cI}
\ee
with $J\neq I$. Eq. \Ref{cI} shows the relation of connections and time derivative of scale factors, as mentioned before.

Now we come to show that $c_Ip_I$ are constants of motion. From Eqs. \Ref{dotp1}-\Ref{dotc2}, we have \label{sec:const}
\be
\f{\rmd}{\rmd t}(c_Ip_I)=-8\pi G\gamma\, \cc.\label{const}
\ee
Thus both of $c_Ip_I$ are constants of motion, since their time derivatives are proportional to the total Hamiltonian constraint

Next, let us introduce the directional Hubble parameters $H_I\equiv \dot{a}_I/a_I=\dot{p}_I/p_I$.
Then using Eqs. \Ref{pa} and \Ref{cI}, the vanishing of Hamiltonian constraint \Ref{hamilton_class} can be written as
 \be
 H_1H_2=8\pi G\rho_{\m},\label{H1H2}
 \ee
 where $\rho_{\m}=p_T^2/2p_1^2p_2^2$ is the energy density of the matter field $T$.
 From Eqs. \Ref{pa} and \Ref{cI}, the directional Hubble parameter can be also related to $c_Ip_I$ by
 \be
 c_1p_1=\gamma V_oa^2H_2,\quad  c_2p_2=\gamma V_oa^2H_1.\label{HI}
 \ee
Here $a\equiv \sqrt{a_1a_2}$ denotes the mean scale factor, which defines the mean Hubble parameter:
\be
H\equiv\f{\dot{a}}{a}=\half(H_1+H_2).\label{hubble}
\ee
Squaring Eq. \Ref{hubble} and using Eq. \Ref{H1H2}, we obtain the generalized Friedmann equation for 2+1 dimensional  Bianchi I models:
\be
H^2=8\pi G\rho_{\m}+\frac{\Sigma^2}{a^4}.\label{friedmann}
\ee
Here
\be
\Sigma^2\equiv \f{a^4}{4}(H_1-H_2)^2\label{shear}
\ee
is the shear term, which can be reexpressed by
 \be
 \Sigma^2=\f{(c_1p_1-c_2p_2)^2}{4\gamma^2  V_o^2},
 \ee
 using Eq. \Ref{HI}. This expression, together with Eq. \Ref{const}, leads to the result that
 $\Sigma^2$ is a constant of motion:
\be
\f{\rmd}{\rmd t}(\Sigma)=0.\label{constSig}
\ee
For the isotropic case, $\Sigma=0$ and Eq. \Ref{friedmann} reduces to the usual Friedmann equation  for 2+1 dimensional  isotropic cosmology \cite{Zhang:2014xqa}.

Now we come to consider the reflections $\Pi_I$:
\be
&\Pi_1(c_1,c_2)=(c_1,-c_2)\\
&\Pi_1(p_1,p_2)=(-p_1,p_2).
\ee
The action of $\Pi_2$ is given by replacement $(1\leftrightarrow2)$. Under each of $\Pi_I$, the Hamiltonian constraint \Ref{hamilton_class} is left invariant. Therefore, in the classical theory,  we can restrict ourselves to the positive octant $p_I\geq 0$, and dynamics in any other octant can be obtained by (combinations of) reflections $\Pi_I$. We will see this reflection $\Pi_I$ will play an important role in quantum theory.

Now we restrict ourselves to the positive octant $p_I\geq 0$, and solve the generalized Friedmann equation \Ref{friedmann}. Using Eqs. \Ref{friedmann} and \Ref{shear}, we have
\be
p_1p_2H_I=\sqrt{4\pi Gp_T^2+\Sigma^2V_o^2}\pm\Sigma V_o.
\ee
Note that $H_I=\dot{p}_I/p_I$, hence we have
\be
p_1p_2=2 t\sqrt{4\pi G p_T^2+\Sigma^2V_o^2}.
\ee
And consequently we have
\be
H_I=\f{\dot{p}_I}{p_I}=\f{1}{2t}\left(1\pm\f{\Sigma V_o}{\sqrt{4\pi G p_T^2+\Sigma^2V_o^2}}\right),
\ee
and its solutions are given by
\be
p_I(t)=p_{I}(0)t^{\kappa_I}\label{pIt}
\ee
where the Kasner exponents $\kappa_I$ are given as \be\kappa_I=\half\pm\f{\Sigma V_o}{\sqrt{16\pi Gp_T^2+4\Sigma^2V_o^2}}.\label{kI}\ee
  From the Hamiltonian constraint \Ref{hamilton_class}, the solution of $T$ is given by
\be
T=T_o+\f{\kappa_T}{\sqrt{8\pi G }}\ln{t},\label{T}
\ee
with
\be
\kappa_T=\f{\sqrt{8\pi G}p_T}{\sqrt{16\pi G p_T^2+4\Sigma^2V_o^2}}\label{kT}.
\ee
 Using \Ref{kI} and \Ref{kT}, we have
 \be
 \kappa_1+\kappa_2=1,\quad\kappa_1^2+\kappa_2^2+\kappa_T^2=1.
 \ee
 The form of the solutions to 2+1 dimensional  Bianchi I model coupled with a massless scalar is very like the one to the 3+1 Bianchi I model \cite{Chiou:2006qq}, as well as the ones to the arbitrary dimensional Bianchi I models \cite{Chen:2000gaa}.
  Combing Eqs. \Ref{pIt} and \Ref{T}, $p_I$ can be given as a function of $T$
\be
p_I(T)=p_I(T_o)e^{\sqrt{8\pi G}\f{\kappa_I}{\kappa_T}(T-T_o)}.
\ee

From the expression of Kasner exponents \Ref{kI}, we have $\kappa_1,\kappa_2>0$. Thus $p_1, p_2$ tend towards zero simultaneously, so do the directional scale factors $a_1,a_2$.\footnote{The approach to singularity for the vacuum 2+1 dimensional  Bianchi I model is different: one of the scale factors tends to zero while the other approaches to constant. } This is different from the singularities in 3+1 Bianchi I model coupled with a massless scalar, where the scale factors in the three directions approach zero not always together, and can be classified into four types \cite{Thorne:1967}.  In 2+1 dimensional  Bianchi I models, the singularity occurs when the energy density of the matter content plays an important role, and the anisotropic shear does not play a dominant role.
\subsection{From the action of 2+1 dimensional  Bianchi I models}
The classical dynamics of 2+1 dimensional  Bianchi I models mentioned here, including the Poisson brackets in Eq. \Ref{cp} and the constraint in Eq. \Ref{hamilton_class}, is reduced from the Hamiltonian framework of full theory of 2+1 dimensional  LQG. Equivalently, it can be also derived by Hamiltonian analysis of the action of Bianchi I models, which is in terms of scale factors. To show this, we start from the action of Bianchi I models:
\be
S_{\gr}=\f{V_o}{8\pi G}\int\rmd t \sqrt{-g} R
\ee
with $\sqrt{-g}=Na_1a_2$  the determinant of the metric given by Eq. \Ref{le}, \be
R=\sum_{I}\left(\f{\ddot{a}_I}{N^2a_I}-\f{\dot{N}\dot{a}_I}{N^3 a_I}\right)+\f{\dot{a}_1\dot{a}_2}{N^2 a_1a_2}
\ee
 the curvature scalar, and $V_o=\int\rmd^2x$ the coordinate volume.  The action in the equation above can be rewritten by
\be
S_{\gr}=-\f{V_o}{8\pi G}\int\rmd t\f{\dot{a}_1\dot{a}_2}{N},
\ee
up to boundary terms. By fixing the coordinate volume $V_o=1$ and the lapse $N=1$, the Lagrangian density is given by
\be
\cl_{\gr}=- \f{\dot{a}_1\dot{a}_2}{8\pi G}.
\ee
The conjugate momentum of $a_I$ is given by
\be
\pi^1:=\f{\p \cl_{\gr}}{\p \dot{a}_1}=-\f{\dot{a}_2}{8\pi G},
\ee
and similarly for $\Pi^2$. The Poisson bracket is given by
\be
\{a_I,\pi^J\}=\delta^J_I.\label{pb_api}
\ee
The Hamiltonian density of gravitational part is given by
\be
\ch_{\gr}=\dot{a}_I\pi^I-\cl_{\gr}=-8\pi G \pi_1\pi_2.
\ee
Minimally coupled with a massless scalar field $T$, the total Hamiltonian density is given by
\be
\ch=-8\pi G \pi_1\pi_2+\f{p^2_{T}}{2a_1a_2}.\label{hamilton_api}
\ee
Using the relations \Ref{pa} and \Ref{cI}, the Poisson brackets \Ref{pb_api} and the Hamiltonian density \Ref{hamilton_api} are equivalent to Eqs. \Ref{cp} and \Ref{hamilton_class}. Thus we can obtain the same generalized Friedmann equation \Ref{friedmann}.

\section{2+1 dimensional  dimensional Loop quantum cosmology of Bianchi I models}\label{sec2}

In the full theory of 2+1 dimensional  LQG, the spectrum of one dimensional ``area" operator is discrete \footnote{In the full theory of 2+1 dimensional  LQG, the spectrum of area operator is discrete, if the gauge group is $SU(2)$, as we consider here. However, if non-compact group $SO(2,1)$ is considered instead,  the spectrum of spacelike intervals is continuous \cite{Freidel:2002hx}. } and has a non-zero minimal value $\Delta$ \cite{Rovelli:1993kc,Thiemann:1997ru,Wisniewski}. We will use this minimal area $\Delta$ to construct the curvature in 2+1 dimensional  LQC. For concreteness,  we
consider a plaquette $\square_{IJ}$ and define curvature along this plaquette. Here the sides of $\square_{IJ}$ are along diagonal directions $I,J$, and the length of the sides are  $\mubar_IL_I$ and $\mubar_JL_J$ respectively measured by fiducial metric ${}^oq_{ab}$; the value of $\mubar$ can be obtained from the minimal area $\Delta$ by some certain scheme.

In $3+1$ loop quantum gravity, the so-called ``improved scheme" leads to very successful quantum dynamics in isotropic case \cite{Ashtekar:2006wn}, where $\mubar\propto1/\sqrt{|p|}$. This scheme is extended to $3+1$  Bianchi I models, in the literature in two different approaches. In Bianchi I models, there are three $p_i$ and three $\mubar_i$ (or $\mubar'$). One scheme assumes that $\mubar_i\propto1/\sqrt{|p_i|}$ \cite{Chiou:2006qq,Chiou:2007sp}, which suffers from fiducial cell scaling and serious problems \cite{Szulc:2008ar,Corichi:2009pp}. The other scheme, where $\mubar'_i\propto1/\sqrt{|a_i|}$ \cite{Bojowald:2007ra,Chiou:2007sp,Ashtekar:2009vc}, solves these problems. Here we will mimic the latter to set $\mubar'_I \propto 1/{|a_I|}$ in 2+1 dimensional  Bianchi I models. The former scheme is also discussed in appendix \ref{alternative}.
\subsection{Quantum Kinematics}
The gravitational part of kinematical Hilbert space is given by $\ch^{\gr}_{\kin}=L^2(\mathbb{R}_{\bohr},\dd\mu_{\bohr})^{\otimes2}$ with the orthonormal basis elements labeled by two real numbers $\ket{p_1,p_2}$, where $\R_{\bohr}$ stands for the Bohr compactification of a real line. The kinematical scalar product is defined as
\be
\bra{p_1,p_2}p'_1,p'_2\rangle=\d_{p_1,p'_1}\d_{p_2,p'_2}.
\ee
Any state $\ket{\Psi}\in\ch^{\gr}_{\kin}$ can be considered as a countable linear combination of this orthonormal basis as
\begin{align}
&\ket{\Psi}=\sum_{p_1,p_2}\Psi(p_1,p_2)\ket{p_1,p_2} \mathrm{\ \ with}\nn \\
&\sum_{p_1,p_2}|\Psi(p_1,p_2)|^2<\infty.\label{norm}
\end{align}

The elemental operators on the gravitational part of kinematical space are momenta $\hat{p}_I$ and   $\SU(2)$ holonomies along edge $e_I$ in the diagonal direction $I$
\be
\widehat{h_I^{(\mubar_I)}}=\widehat{\cos\f{c_I\mubar_I}{2}}\mathbb{I}+2\widehat{\sin\f{c_I\mubar_I}{2}}\tau_I,
\label{holonomy}
\ee
where $\mathbb{I}$ is the identity matrix,  $\tau_I=\sigma_I/2i$ with $\sigma_i$ the Pauli matrices, and $\mubar_IL_I$ is the length of the edge with respect to ${}^oq_{ab}$. The action of the elemental operators on the gravitational part of kinematical space $\ch^{\gr}_{\kin}$ are given by
\begin{align}
\hat{p}_I \Psi(p_1,p_2)&=p_I\Psi(p_1,p_2),\\
\widehat{\exp{(i\mubar_1c_1)}}\Psi(p_1,p_2)&=\Psi(p_1-8\pi G\gamma\hbar \mubar_1,p_2),
\end{align}
and similarly for $\widehat{\exp{(i\mubar_2c_2)}}$.

Next, for the matter part, the scalar field $T$ is quantized as usual:
\begin{align}
\hat{T}\Psi(p_1,p_2,T)&=T\Psi(p_1,p_2,T),\nn\\\hat{p}_T\Psi(p_1,p_2,T)&=-i\hbar\f{\rmd}{\rmd T}\Psi(p_1,p_2,T),\nn
\end{align}
where $\Psi(p_1,p_2,T)\in \ch^{\gr}_{\kin}\otimes L^2(\R,\rmd T)$.
In the next subsection, we will introduce the Hamiltonian operator in terms of these elemental operators, acting on the total kinematical space $\ch_{\kin}=\ch^{\gr}_{\kin}\otimes L^2(\R,\rmd T)$.

\subsection{Construction of Hamiltonian operator}
To construct the Hamiltonian constraint operator, we first express the curvature $F_{ab}{}^k$ in terms of holonomies:
\be
F_{ab}{}^k=-2\sum_{I,J}\tr\left(\f{h_{\square_{IJ}-\mathbb{I}}}{\mathrm{Ar}_{IJ}}\tau^k\right){}^o\omega_a^I{}^o\omega_b^J.
\label{Fab}
\ee
Here $h_{\square_{IJ}}\equiv h^{(\mubarp_I)}_Ih^{(\mubarp_J)}_Jh^{(\mubarp_I)-1}_Ih^{(\mubarp_J)-1}_J$ is the holonomy around the plaquette $\square_{IJ}$, whose sides are along diagonal directions $I,J$ and have length $\mubarp_IL_I$ and $\mubarp_JL_J$ respectively measured by fiducial metric ${}^oq_{ab}$; $\mathrm{Ar}_{IJ}=\mubarp_I\mubarp_J$ is the area of the plaquette $\square_{IJ}$.
And $\mubarp_I$ is given as
\be
\mubarp_1=\frac{\Delta}{|p_2|}, \quad \mubarp_2=\frac{\Delta}{|p_1|},\label{mubar}
\ee
where $\Delta$ is the minimal 1-dimensional ``area''.
Using the formula \Ref{Fab} of curvature together with Thiemann's trick in 2+1 dimensional  LQG \cite{Thiemann:1997ru}
\be
\half \epsilon^{ijk}\epsilon_{ab}E^a_jE^b_k=\f{1}{2(8\pi G\gamma)^2}\epsilon_{ijk}\epsilon^{ab}\{A_a^j,V\}\{A_b^k,V\},\nn
\ee
the operator of the gravitational part of the Hamiltonian constraint
\be
\cc_{\gr}=-\f{1}{16\pi G\gamma^2}\int_{\cv}\f{E^a_iE^b_j}{\sqrt{q}}\epsilon^{ij}{}_{k}F_{ab}{}^k
\ee
can be given by
\bw
\begin{align}
 {\cc}_{\mathrm{gr}}=&\pam
\sum_{I,J,K,L}\f{\eps^{IJ}\eps^{KL}}{\mubarp_I\mubarp_J\mubarp_K\mubarp_L}\tr(h_{\square_{IJ}}h_K\{h_K^{-1},\sqrt{V}\}h_L\{h_L^{-1},\sqrt{V}\}),
\end{align}
where $h_K$ is short for $h^{(\mubarp_K)}_K$, the holonomy along edge $e_K$ in the diagonal direction $K$. If the Poisson brackets are replaced by commutators : $\{,\}\rightarrow [,]/i \hbar$, and if the observables are replaced by the corresponding operators, we will obtain the operator of the gravitational part of the hamiltonian constraint, up to factor orderings
\begin{align}
\hat{\cc}_{\gr}=&\pam\f{1}{\hbar^2}\sum_{I\neq J,K\neq L}\sin(\mubarp_Ic_I)\sin(\mubarp_Jc_J)\f{\mubarp_K\mubarp_L}{\mubarp_I\mubarp_J}\hat{A}_{K}\hat{A}_{L}
\end{align}
where
\be
\hat{A}_{K}=(\mubarp_K)^{-2}\left(\cos(\f{\mubarp_K c_K}{2})\sqrt{V}\sin(\f{\mubarp_K c_K}{2})-\sin(\f{\mubarp_K c_K}{2})\sqrt{V}\cos(\f{\mubarp_K c_K}{2})\right).
\label{f}
\ee
\ew
By expanding the summation over $I\neq J, K\neq L$, we will find four terms of the same value.
Thus
\begin{align}
\hat{\cc}_{\mathrm{gr}}=\f{8}{(8\pi G)^3\gamma^4\hbar^2} \sin(\mubarp_1c_1)\sin(\mubarp_2c_2)A_1A_2.
\end{align}
The symmetrized operator is given by
\begin{align}
\hat{\cc}_{\gr}=  \sin((\mubarp_1c_1))\hat{A}\sin(\mubarp_2c_2)+\sin(\mubarp_2c_2)\hat{A}\sin(\mubarp_1c_1),\nn\\
\label{hamilton_sym}
\end{align}
with
\begin{align}
\hat{A}=\f{4}{(8\pi G)^3\gamma^4\hbar^2} \hat{A}_1\hat{A}_2.\label{F}
\end{align}

Before going to study the action of the Hamiltonian operator in next subsection, we show the induced action of reflections $\Pi_I$ on the classical phase space.

On the space of wave functions $\Psi(p_1,p_2)$, the two reflections $\Pi_I$ on the classical phase space  have a natural induced action $\hat{\Pi}_I$, for example, $\hat{\Pi}_1\Psi(p_1,p_2)=\Psi(-p_1,p_2)$. We will assume that the wave function $\Psi(p _1,p_2)$ is symmetric under the action $\hat{\Pi}_I$, which implies
\be
\Psi(p_1,p_2)=\Psi(|p_1|,|p_2|).
\ee
On the quantum operators $\hat{\co}$, the induced action of reflections $\hat{\Pi}_I$ is given by
\be
\hat{\co}\rightarrow(\hat{\Pi}_I\hat{\co}\hat{\Pi}_I)\Psi:=\hat{\Pi}_I\hat{\co}\hat{\Pi}_I\Psi.
\ee
Action of the reflections $\hat{\Pi}_I$ on elemental operators is given by
\begin{align}
&\hat{\Pi}_I\hat{p}_J\hat{\Pi}_I=s_{_{IJ}}\hat{p}_J\label{pi_p}\\
&\hat{\Pi}_I\widehat{\exp{(\pm i\mubarp_Jc_J)}}\hat{\Pi}_I=\widehat{\exp{(\pm s_{_{IJ}} i\mubarp_Jc_J)}}\label{pi_c}
\end{align}
with $s_{_{IJ}}=\pm1$ given by
\begin{equation}
  s_{_{IJ}}=\left\{
   \begin{array}{l}
   -1\ \mathrm{if}\ I=J; \\
   1\ \mathrm{if}\ I\neq J. \end{array}
   \right.
  \end{equation}
  For the matter part, both of $\hat{T}$ and $\hat{p_T}$ are invariant under reflections $\hat{\Pi}_I$.
From the action of reflections $\hat{\Pi}_I$ on the elemental operators \Ref{pi_p} and \Ref{pi_c}, we find the gravitational part of Hamiltonian operator \Ref{hamilton_sym} is reflection symmetric:
\be
\hat{\Pi}_I\hat{\cc}_{\gr}\hat{\Pi}_I=\hat{\cc}_{\gr},
\ee
just as in the classical theory. Therefore,  its action is well defined on $\ch_{\kin}^{\gr}$.

\section{Action of Hamiltonian operator}\label{sec3}
To see explicitly the action of the Hamiltonian operator \Ref{hamilton_sym}, we introduce a new orthonormal basis $\ket{\l_1,\l_2}$ in $\Hil^{\gr}_{\ki}$ with
\be
\l_I\equiv \frac{p_I}{\sqrt{8\pi|\gamma|\de\lp}},
\ee
where $\lp=G\hbar$ is the Planck length in 2+1 dimensions .
The action of elemental operators on wave functions $\Psi(\l_1,\l_2)$ is given by
\begin{align}
&(p_I\Psi)(\l_1,\l_2)=\sqrt{8\pi|\gamma|\de\lp}\l_I\Psi(\l_1,\l_2),\\
&\exp (\pm i\mubarp_1c_1)\Psi(\l_1,\l_2)=\Psi\left(\l_1\mp \f{1}{|\l_2|},\l_2\right),\label{exp_lambda}
\end{align}
and similar for $\exp (\pm i\mubarp_2 c_2)$. Equation \Ref{exp_lambda} can be given by the expression \Ref{mubar} of $\mubarp_I$ and the Poisson brackets, which imply
\begin{align}
\exp (\pm i\mubarp_Ic_I)&=\exp \left(\mp \f{8\pi\gamma\de\lp}{|p_J|}\f{\rmd}{\rmd p_I} \right),\\
&=\exp \left(\mp \f{1}{|\l_J|}\f{\rmd}{\rmd \l_I} \right)=:E^{\mp}_I
\end{align}
with $J\neq I$.

To make the quantum dynamics easier to compare with that of the Friedmann models in 2+1 dimensions  \cite{Zhang:2014xqa}, we introduce the volume of the elementary cell $\mathcal{V}$ as one of the arguments of the wave function, mimicking \cite{Ashtekar:2009vc}.  Set
\be
v=2\l_1\l_2,
\ee
and use $\l_1, v$ as the configuration variables in place of $\l_1, \l_2$, then the action of the volume of $\mathcal{V}$ on $\Psi(\l_1,v)$ is given by
\be
\hat{V}\Psi(\l_1,v)=4\pi|\gamma|\de\lp |v| \Psi(\l_1,v).
\ee

We now restrict the argument of $\hat{\cc}_{\gr}$ to the positive octant.
The action of $\hat{A}_I$ in equation \Ref{f} on $\Psi(\l_1,v)$ is given by
\be
\hat{A}_I\Psi(\l_1,v)=\f{4\pi|\gamma|\lp}{i\de}\l^2_J(\sqrt{|v-1|}-\sqrt{|v+1|})\Psi(\l_1,v),\nn
\ee
with $J\neq I$. Then the action of $\hat{A}$ in equation \Ref{F} on $\Psi(\l_1,v)$ is given by
\be
\hat{A}\Psi(\l_1,v)=-\f{\pi\lp}{8 \pi G\gamma\de}v^2(\sqrt{|v-1|}-\sqrt{|v+1|})^2\Psi(\l_1,v).\nn
\ee
Now let us give the action of the gravitational part \Ref{hamilton_sym} of the Hamiltonian operator:
\begin{align}
\hat{\cc}_{\gr}\Psi(\l_1,v)&=f(v+2)(\Psi^+_4-\Psi^+_0)-f(v-2)(\Psi^-_0+\Psi^-_4),\label{hamilton_operator}
\end{align}
where
\be
f(v)=\f{\pi\lp}{32 \pi G\gamma\de}v^2(\sqrt{|v-1|}-\sqrt{|v+1|})^2,\nn
\ee
and $\Psi^{\pm}_{0,4}$ are defined as follows:
\begin{align}
\Psi^{\pm}_4&=\Psi(\f{v\pm 2}{v}\l_1,v\pm 4)+\Psi(\f{v\pm 4}{v\pm 2}\l_1,v\pm 4),\\
\Psi^{\pm}_0&=\Psi(\f{v\pm 2}{v}\l_1,v)+\Psi(\f{v }{v\pm 2}\l_1,v ).
\end{align}

Now we come to consider the operator of the matter part of the Hamiltonian constraint, whose  classical expression is given by
\be
\cc_{\m}=\f{p_{T}^2}{2V}.
\ee
When the inverse volume operator corresponding to $1/V$ is defined, three ambiguities appear \cite{Singh:2013ava}:
\be
\widehat{1/V}=(\hat{V})^p\left(\f{1}{2s\alpha}\left||\hat{V}+\alpha|^s-|\hat{V}-\alpha|^s\right|\right)^{\f{1+p}{1-s}},
\ee
with $p>0,\ \alpha>0,$  and $0<s<1$. These are all reasonable inverse volume operators, since  they annihilate zero volume states and approximate to $1/V$ for large $V$.
The particular inverse volume operator in 2+1 dimensional  FLRW model \cite{Zhang:2014xqa} is obtained by setting $\alpha=4\pi\gamma\de\lp,\ s=\f{1}{4}$ and $p=2$:
\begin{align}
\widehat{V^{-1}}\Psi(\l_1,v)&=
\f{4|v|^2}{\pi\gamma\de\lp}\left|{|v+1|^{{1}/{4}}}-{|v-1|^{{1}/{4}}}\right|^4\Psi(\l_1,v)\nn\\
&=\f{B(v)}{4\pi\gamma\de\lp}\Psi(\l_1,v),
\end{align}
where $B(v)\equiv 16 \left|{|v+1|^{{1}/{4}}}-{|v-1|^{{1}/{4}}}\right|^4$ is different from the one defined in 2+1 dimensional  FLRW model \cite{Zhang:2014xqa} by a factor of $4\pi\gamma\de\lp$.
We will use this particular inverse volume operator in the following, to have the Hamiltonian operator comparable to the one of 2+1 dimensional  FLRW model. Thus the action of the matter part of the Hamiltonian operator on wave functions $\Psi(\l_1,v;T)\in\ch_{\kin}=\ch^{\gr}_{kin}\otimes L^2(\R,\rmd T)$ is given by
\be
\hat{\cc}_{\m}\Psi(\l_1,v;T)=-\f{\hbar^2}{4\pi\gamma\de\lp} B(v){\p^2_T}\Psi(\l_1,v;T).\label{hamilton_matter}
\ee
Collecting the gravitational part \Ref{hamilton_operator} and the matter part \Ref{hamilton_matter} of the Hamiltonian operator, we can express the vanishing of the total Hamiltonian operator as
\bw
\be
{\p^2_T}\Psi(\l_1,v;T)=[B(v)]^{-1}\left(C_+(v)\left(\Psi^+_4(T)-\Psi^+_0(T)\right)-C_-(v)\left(\Psi^-_0(T)+\Psi^-_4(T)\right)\right),
\label{hamilton_diff}
\ee
\ew
where
\begin{align}
C_+(v)=&\f{4\pi\gamma\de\lp}{\hbar^2}f(v+2)\nn\\
=&\f{\pi}{8\hbar}(v+2)^2(\sqrt{|v+1|}-\sqrt{|v+3|})^2,\\
C_-(v)=&C_+(v-4)
\end{align}
and $\Psi^{\pm}_{0,4}(T)$ are defined as follows:
\begin{align}
\Psi^{\pm}_4(T)&=\Psi(\f{v\pm 2}{v}\l_1,v\pm 4;T)+\Psi(\f{v\pm 4}{v\pm 2}\l_1,v\pm 4;T),\\
\Psi^{\pm}_0(T)&=\Psi(\f{v\pm 2}{v}\l_1,v;T)+\Psi(\f{v }{v\pm 2}\l_1,v;T ).
\end{align}
\section{Properties of the quantum dynamics}\label{sec4}
\subsection{Relation to the 2+1 dimensional  LQC Friedmann dynamics}
In classical theory, the Friedmann model can be reduced from the Bianchi I model by applying isotropic conditions $a_1=a_2$ in 2+1 dimensional case or $a_1=a_2=a_3$ in 3+1 dimensional case. In 3+1 LQC, it is shown in \cite{Ashtekar:2009vc} that
there is a natural projection from a dense subspace of the physical Hilbert space of Bianchi I model to that of the Friedmann model. In this subsection, we will mimic the 3+1 projection to construct a 2+1 dimensional  projection, which maps the Bianchi I Hamiltonian constraint \Ref{hamilton_diff} to that of 2+1 dimensional  Friedmann model.
The idea is to integrate out the extra, anisotropic degrees of freedom, which first appear in \cite{Bojowald:2005wh}.

We define a projection $\mathbb{P}$ from states $\Psi(\l_1,v)$ of the 2+1 dimensional  Bianchi I models to the states $\psi(v)$ of 2+1 dimensional  Friedmann model of \cite{Zhang:2014xqa} as follows:
\be
\Psi(\l_1,v)\rightarrow(\mathbb{P}\Psi)(v)=\sum_{\l_1}\Psi(\l_1,v)\equiv \psi(v).
\ee
Applying this projection map to the Hamiltonian operator \Ref{hamilton_diff}, we find
\bw
\be
{\p^2_T}\psi(v;T)=2[B(v)]^{-1}\left(C_+(v)\left(\psi(v+4;T)-\psi(v;T)\right)-C_-(v)\left(\psi(v-4;T)+\psi(v;T)\right)\right),
\ee
\ew
which is the total Hamiltonian operator of 2+1 dimensional  Friedmann model. This result shows there is a simple and exact relation between quantum dynamics of Bianchi I model and Friedmann model.
\subsection{Effective equations}
Because of the complexity of Bianchi I model, and  the set of our $\mubarp_I$, it is not easy to carry out the semi-classical analysis and derive the effective equation. In this subsection, we obtain effective equation by analog of the classical equation, replacing $c_I$ by $\sin{(\mubarp_Ic_I)}/\mubarp_I$. The effective equation obtained in this approach shows to be the same with the one derived by semi-classical analysis in isotropic case \cite{Ding:2008tq,Zhang:2014xqa}
and the $\mubarp_i\propto1/\sqrt{|p_i|}$ scheme of Bianchi I model \cite{Chiou:2006qq}.

The effective Hamiltonian constraint is given by the analog of classical form \Ref{hamilton_class}:
\be
\cc^{\eff}=\cc_{\gr}^{\eff}+p_1p_2\rho_{\m},\label{hamilton_eff}
\ee
where
\be
\cc^{\eff}_{\gr}=-\frac{p_1p_2}{8\pi G\gamma^2\de^2}\sin{(\mubarp_1c_1)}\sin{(\mubarp_2c_2)}.
\ee
The vanishing of the effective Hamiltonian constraint \Ref{hamilton_eff} gives upper bound of the matter density:
 \be
 \rho_{\m}=\rho_{c}\sin{(\mubarp_1c_1)}\sin{(\mubarp_2c_2)}\leq\rho_{c},\label{rhoc}
 \ee
 where $\rho_{c}=1/8\pi G\gamma^2\de^2$ is the maximal density.
 Note that the matter density becomes infinite at the big-bang singularity in the classical evolution, thus the upper-bounded density shows that the singularity is resolved in the effective theory. This is different from the case in 3+1 dimensions. Remind that in 3+1 dimensions, the planar collapse for
Kasner-like solutions is not resolved by the quantum effect and thus the vanishing behavior of the length scale factors $a_I$ remains \cite{Chiou:2006qq}. In 2+1 dimensions, we only have one type of solution and  the spectrum of length operator is discreet, thus all the singularity appeared in 2+1  dimensional classical Bianchi I model is resolved.

Using Eq. \Ref{rhoc}, we can go further to  define the effective  ``directional" matter density,
 \be
 \rho_I=\rho_c \sin^2{(\mubarp_Ic_I)},\label{rhoI}
 \ee
 which is also bounded by the  maximal density $\rho_{c}$ and
 \be \rho_1\rho_2=\rho_{\m}^2.\ee

Effective equations are obtained via Poisson brackets
\begin{align}
\dot{p}_1=&\{p_1,\cc^{\eff}\}=\f{p_1}{\gamma\de}\cos{(\mubarp_1c_1)}\sin{(\mubarp_2c_2)},\\
\dot{c}_1=&\f{c_1p_2}{\gamma\Delta p_1}\sin{(\mubarp_1c_1)}\cos{(\mubarp_2c_2)}-16\pi G\gamma p_2\rho_{\m}\nn\\
=&\f{c_1}{p_1}\dot{p}_2-16\pi G\gamma p_2\rho_{\m},\end{align}
where the vanishing of effective Hamiltonian constraint \Ref{hamilton_eff} is used, and $\dot{p}_2$, $\dot{c}_2$ can be obtained by replacement of $(1\leftrightarrow2)$.

 The effective  directional Hubble parameter is thus given by
\be
H_I=\f{\dot{p}_I}{p_I}=\f{1}{\gamma\de}\cos{(\mubarp_Ic_I)}\sin{(\mubarp_Jc_J)},
\label{effHI}
\ee
with $J\neq I$.
Using definition of the shear in Eq. \Ref{shear} and the effective directional Hubble parameter in Eq. \Ref{effHI}, the effective shear is given by
\be
\bar{\Sigma}=-\f{p_1p_2}{2\gamma\de V_o}\sin{(\mubarp_1c_1-\mubarp_2c_2)},\label{effshear}
\ee
which implies the fact that the effective shear is finite throughout the evolution. However, in effective theory, the shear is no longer a constant of motion as it is in classical theory.

Using the effective directional Hubble parameter in Eq. \Ref{effHI}, the effective mean Hubble parameter can be given by
\be
H=\half(H_1+H_2)=\f{1}{2\gamma\de}
\sin{(\mubarp_1c_1+\mubarp_2c_2)}.\label{effH}
\ee
Using Eq.s \Ref{rhoc}, \Ref{rhoI} and \Ref{effshear}, the square of the effective Hubble parameter in Eq. \Ref{effH} can be given by
\be
H^2=8\pi G\rho_{\m}\sqrt{1-\f{\rho_1}{\rho_c}}\sqrt{1-\f{\rho_2}{\rho_c}}+\frac{\bar{\Sigma}^2}{a^4}.
\label{effriedmann}\ee
For the isotropic case, the effective shear term vanishes, and $\rho_1=\rho_2=\sqrt{\rho_{\m}}$, thus the effective generalized Friedmann equation \Ref{effriedmann} will reduce to the usual Friedmann equation in 2+1 dimensional  Friedmann model \cite{Zhang:2014xqa}: $H^2=8\pi G\rho_{\m}(1-{\rho_{\m}}/{\rho_c})$. For the classical limit, $\mubarp_Ic_I<<1$, we have $\sin{(\mubarp_Ic_I)}\rightarrow \mubarp_Ic_I$ and $\cos{(\mubarp_Ic_I)}\rightarrow 1$, thus the effective generalized Friedmann equation in \Ref{effriedmann} will approximately become the classical generalized Friedmann equation \Ref{friedmann}. If we drop the higher order, Eq. \Ref{effriedmann} becomes
\be\label{effriedmann2}
H^2=8\pi G\rho_{\m}\left(1-\f{\rho_{\m}}{\rho_c}\right)+
\frac{\bar{\Sigma}^2}{a^4}\left(1-\f{2\rho_{\m}}{\rho_c}\right)\nn\\+\co\left((\mubarp_Ic_I)^4\right).
\ee
Again we can see in the classical limit $\rho_{\m}<<\rho_c$, it will go to the generalized Friedmann equation \Ref{friedmann}. Vanishing of the Hubble parameter in Eq. \Ref{effriedmann2} will give the matter energy density at the
bounce
\be
\rho_{\bounce}\approx\f{\rho_c}{2}-\f{\sigbar^2}{8\pi Ga^4}+
\sqrt{\left(\f{\rho_c}{2}\right)^2+\left(\f{\sigbar^2}{8\pi G a^4}\right)^2},\nn
\ee
which is bounded by $\rho_c$.

\section{Conclusions}
This paper gives the detailed construction of loop quantum cosmology of Bianchi I models in 2+1 dimensions . Both the $\mubar$ and $\mubar'$ schemes which appeared in four dimensional case are successfully deployed. In order to make this paper as compact as possible, we only discuss the more physical intuitively  $\mubar'$ scheme in the main text, while leave the discussion of $\mubar$ scheme in the appendix. Our results show that the discreteness of the underlying quantum geometry of 2+1 dimensions  again gives rise to a difference equation which represents the evolution of the three dimensional universe.

Meanwhile, we interestingly observe that in 2+1 dimensions  also admitting some new features. More precisely, in 3+1 dimensions, the planar collapse for
Kasner-like solutions is not resolved by the quantum effect and thus the vanishing behavior of the length scale factors $a_I$ remains. However, we only have one type of solution and note that in 2+1 dimensional case, the spectrum of length operator is discreet, thus all the singularity appeared in 2+1 dimensional classical Bianchi I model is resolved. Of course, the results in present paper is still quite preliminary and requires further investigations.

There are several possible extensions of our results. The first one is to generalize our result to higher dimensions. while the second one is more interesting and subtly. Namely, our results can serve as the first step of link LQC from LQG in 2+1 dimensions.  Indeed, some efforts have been done towards this important direction directly from 3+1 dimensions \cite{Alesci:2012md,Alesci:2013xd,Alesci:2013xya,Alesci:2014uha,Alesci:2014rra,Alesci:2015jca,Alesci:2015nja,Bilski:2015dra,
Bodendorfer:2014wea,Bodendorfer:2015aca,Bodendorfer:2014vea,Bodendorfer:2015hwl}. However, since the complete understand of the dynamics of LQG in 3+1 dimensions is still lacking. Therefore lessons from 2+1 dimensions  might be valuable, since the 2+1 dimensional quantum gravity is proven to be exactly solveable.

\section*{Acknowledgments}

This work is supported by NSFC with No.11305063  and No. 11205013,
the Fundamental Research Funds for the Central University of China, and SRFDP with No. 20120009120039.

\appendix
\section{Alternate quantization}\label{alternative}
In this appendix, we consider the effective dynamics from an alternative scheme, which assumes that $\mubar_i\propto1/\sqrt{|p_i|}$  in 3+1 Bianchi I model \cite{Chiou:2006qq,Chiou:2007sp}. For 2+1 dimensional  Bianchi I model, $\mubar$ in this scheme is given as
\be
\mubar_I=\frac{\Delta}{|p_I|}.\label{mubar_chiou}
\ee
In this scheme, the effective Hamiltonian constraint is given by
\be
\cc^{\eff}=\cc_{\gr}^{\eff}+p_1p_2\rho_{\m},\label{hamilton_eff_chiou}
\ee
where
\be
\cc^{\eff}_{\gr}=-\frac{p_1p_2}{8\pi G\gamma^2\de^2}\sin{(\mubar_1c_1)}\sin{(\mubar_2c_2)}.
\ee
Although Eq. \Ref{hamilton_eff_chiou} seems the same in form as Eq. \Ref{hamilton_class}, they are different, since they have different definition of $\mubar_I$. Since they are same in form, the vanishing of Eq. \Ref{hamilton_eff_chiou} gives the same upper bound of the matter density:
\be
 \rho_{\m}=\rho_{c}\sin{(\mubar_1c_1)}\sin{(\mubar_2c_2)}\leq\rho_{c}.
 \ee
 The effective equations are different:
 \begin{align}\label{dotp1_chiou}
\dot{p}_1=&\f{p_2}{\gamma\de}\cos{(\mubar_1c_1)}\sin{(\mubar_2c_2)}\\
\dot{c}_1=&\f{c_1p_2}{\gamma\Delta p_1}\cos{(\mubar_1c_1)}\sin{(\mubar_2c_2)}-16\pi G\gamma p_2\rho_{\m}\nn\\
=&\f{c_1}{p_1}\dot{p}_1-16\pi G\gamma p_2\rho_{\m},\end{align}
where the vanishing of the effective Hamiltonian constraint \Ref{hamilton_eff_chiou} is used. We have shown in section \ref{sec:const} that $c_Ip_I$ are constants of motion in classical theory. Their effective analogs in the scheme \Ref{mubar_chiou} can be defined as
\be\label{GI}
\cg_I:=p_I\f{\sin{(\mubar_Ic_I)}}{\mubar_I}=\f{p_I^2}{\Delta}\sin{(\mubar_Ic_I)},
\ee
with the time derivative
\be
\f{\rmd }{\rmd t}\cg_{I}=0.
\ee
Hence $\cg_I$ are constants of motion in effective theory. However, this form defined in the scheme \Ref{mubar}  is not constant anymore.
 The vanishing of the effective Hamiltonian constraint \Ref{hamilton_eff_chiou} gives
 \be
 \cg_1\cg_2=4\pi G\gamma^2p_T^2.
 \ee

 Using Eqs. \Ref{dotp1_chiou} and \Ref{GI}, we have
 \be
 \f{\rmd p_1}{\rmd T}=\f{\cg_2}{\gamma p_T}\sqrt{p_1^2-\f{\cg_1^2\Delta^2}{p_1^2}},
 \ee
 which gives the solution to the effective equations:
 \be
 p_1(T)=\sqrt{\half\left(\cg_1^2\Delta^2e^{-\f{2\cg_2}{\gamma p_T}(T-T_o)}+e^{\f{2\cg_2}{\gamma p_T}(T-T_o)}\right)},
 \ee
 and similarly for $p_2(T)$.

\def\cprime{$'$}

\end{document}